\documentclass[12pt]{iopart}
\usepackage{iopams}
\usepackage{graphicx}
\usepackage{dsfont}
\usepackage{pxfonts}

\newtheorem{prop}{Proposition}

\newcommand{\half}{\mbox{$\textstyle \frac{1}{2}$}} 
\newcommand{\quat}{\mbox{$\textstyle \frac{1}{4}$}} 
\newcommand{\re}{{\rm e}}
\newcommand{\ri}{{\rm i}}
\newcommand{\rd}{{\rm d}}

\begin{document}

\title[Biorthogonal Quantum Mechanics]
{Biorthogonal Quantum Mechanics}

\author[D.~C.~Brody]{Dorje C. Brody}%

\address{Mathematical Sciences, Brunel University, Uxbridge UB8 3PH, UK}%

\begin{abstract}
The Hermiticity condition in quantum mechanics required for the characterisation of (a) physical observables and (b) generators of unitary motions can be relaxed into a wider class of operators whose eigenvalues are real and whose eigenstates are complete. In this case, the orthogonality of eigenstates is replaced by the notion of biorthogonality that defines the relation between the Hilbert space of states and its dual space. The resulting quantum theory, which might appropriately be called 'biorthogonal quantum mechanics', is developed here in some detail in the case for which the Hilbert space dimensionality is finite. Specifically, characterisations of probability assignment rules, observable properties, pure and mixed states, spin particles, measurements, combined systems and entanglements, perturbations, and dynamical aspects of the theory are developed. The paper concludes with a brief discussion on infinite-dimensional systems.  
\end{abstract}

\submitto{\JPA}

\section{Introduction}
\label{sec:1}

In standard quantum mechanics observable quantities are characterised by Hermitian operators. 
The eigenvalues of a Hermitian operator represent possible outcomes of the measurement of an 
observable represented by that operator. Once the 
measurement of, say, the energy is performed and the outcome recorded, the system is in a state 
of definite energy, that is, there cannot be a transition into another state with a different energy. 
Hermitian operators conveniently encode this feature in the form of the orthogonality of their 
eigenstates. 

The observed lack of transition into another state, however, can only be translated into the 
abstract `mathematical' notion of the orthogonality of states in Hilbert space via the specification 
of the 
probability rules in quantum mechanics. When eigenstates of an observable are not orthogonal, 
however, there is an equally natural way of assigning probability rules so that the resulting quantum 
theory appears identical to the conventional theory. Evidently, in this case observables are not 
represented by conventional Hermitian operators, since otherwise the eigenstates are necessarily 
orthogonal. Nevertheless, if an operator has a complete set of eigenstates and real eigenvalues, 
then it becomes a viable candidate for representing a physical observable. The key mathematical 
ingredients required to represent physical observables are that the eigenvalues are real, and that 
eigenstates are complete; whereas the notion of orthogonality can be relaxed and substituted by a 
weaker requirement of biorthogonality. The resulting quantum theory will thus be called biorthogonal 
quantum mechanics. 

There is a substantial literature on the idea of relaxing the Hermiticity requirement for 
observables in quantum mechanics. For example, Scholtz \textit{et al}. \cite{SGH,GHS} proposes 
the introduction of a nontrivial metric operator in Hilbert space and defines physical observables as 
self-adjoint operators with respect to the choice of the metric. Viewed from the conventional `flat' 
inner-product structure, therefore, observables are no longer Hermitian and their eigenstates are 
not orthogonal, but in the Hilbert space endowed with this nontrivial metric we recover the `standard' 
quantum theory. Bender and others have developed PT-symmetric quantum theory where the 
Hermiticity condition is replaced by the invariance under simultaneous parity and time reversal 
operation. A PT-symmetric Hamiltonian is in general not Hermitian, but if the corresponding 
eigenstates are also PT symmetric, then the eigenvalues are real and eigenstates may be 
complete, and can be used to describe quantum systems \cite{CMB0,mostafa1,znojil}. 
Operators that are not Hermitian also play an important 
role in the physics of resonance, as discussed, for example, in \cite{Moiseyev,IR,Moiseyev2}. The 
role of biorthogonal systems in PT-symmetric quantum theories is discussed in 
Curtright \& Mezincescu \cite{Curtright}. 

The works mentioned here are detailed and substantial, and contain a large number of 
references. In spite of this, here we shall present `yet another account' of the subject since a 
number of basic and foundational ideas of quantum mechanics, already required for the 
representation of quantum systems modelled on finite-dimensional Hilbert spaces, such as a 
detailed account of probabilistic interpretations, a characterisation of measurement processes, 
or a formulation of combined systems and the role of entanglements, have not been made 
completely transparent. It turns out that the approach based from the outset on the use of 
biorthogonal basis (as in \cite{Curtright}) allows us to develop these basic ideas in the most 
elementary manner. The purpose of the present paper therefore is to develop the formalism 
of biorthogonal quantum mechanics for systems modelled on finite-dimensional Hilbert spaces, 
and along the way clarify various issues in a transparent and accessible way. 

The paper will be organised as follows. We begin in \S\ref{sec:2} with an overview of the 
biorthogonal system of basis in Hilbert space that arise from the eigenstates of a complex 
(i.e. not necessarily Hermitian) Hamiltonian and those of its Hermitian adjoint, for the benefit 
of readers less acquainted with the material. The effectiveness of the use of biorthogonal basis 
associated with operators that are not self adjoint has a long history and goes back to the work 
of Liouville \cite{Liouville}, subsequently developed further by Birkhoff \cite{Birkhoff}. In the case 
of a real Hilbert space of square-integrable functions defined on a finite interval of the real line 
${\mathds R}$, properties of biorthogonal bases associated with operators that are not self 
adjoint have been worked out in detail by Pell \cite{Pell,Pell2}. Many of the results, with suitable 
modifications, extend into the complex domain, as developed by Bari \cite{Bari} (cf. \cite{gelfand}). 

In \S\ref{sec:3} we establish the relation between the Hilbert space ${\mathcal H}$ of states and 
its dual space ${\mathcal H}^*$, and this in turn leads to the identification of a consistent 
probability assignment for transitions between states. It will be shown that although eigenstates 
of a complex Hamiltonian are not orthogonal in ${\mathcal H}$, they nevertheless do correspond 
to maximally separated states in the ray-space, hence there cannot be transitions between these 
states. An analogous conclusion has been drawn previously (e.g., in \cite{mostafa1}), 
but it will become evident that the biorthogonal method employed here leads to this result in the 
most elementary fashion, without referring to heavy-handed mathematical arguments. The 
construction of observables, their expectations, as well as the notion of general mixed states, are 
then developed in some detail in \S\ref{sec:4}. 

In \S\ref{sec:5} we discuss measurement-theoretic and further probabilistic aspects of complex 
Hamiltonians. 
It will be shown, in particular, that for unitary systems 
orthogonality of eigenstates in ${\mathcal H}$ is not a condition 
that can be asserted from experiments, thus making any operator having a complete set of 
eigenstates and real eigenvalues a viable candidate for the representation of observable quantities. 
The construction of combined systems in biorthogonal quantum mechanics is then developed in 
\S\ref{sec:6}, where we also define coherent states in this context. 
In \S\ref{sec:7} we describe how the Rayleigh-Schr\"odinger perturbation theory works in the case 
of complex Hamiltonians. Perturbation of complex Hamiltonians away from eigenstate degeneracies 
in fact has been known for some time \cite{more,SW}. The purpose of this section is to give a brief 
review of the idea, partly for completeness and partly on account of the fact that the result provides 
an independent confirmation that the probability assignment rule of \S\ref{sec:3} is in some sense 
the `correct' one. Properties of time evolution of quantum states generated by a complex 
Hamiltonian 
are described in \S\ref{sec:8}, showing that reality and completeness lead to unitarity, without 
the orthogonality requirement. 
In \S\ref{sec:9} we turn to the discussion of PT-symmetric quantum 
mechanics, in particular how it ties in with the notion of biorthogonal quantum mechanics. We conclude in 
\S\ref{sec:10} with a brief discussion towards subtleties arising from the consideration of quantum 
systems described by infinite-dimensional Hilbert spaces. 



\section{Eigenstates of complex Hamiltonians and their adjoints}
\label{sec:2}

To begin the analysis of quantum mechanics using basis functions that are in general not 
orthogonal, we shall first review basic properties of eigenstates of generic complex 
Hamiltonians in finite dimensions. Let ${\hat K}={\hat H}-\ri {\hat{\mathit\Gamma}}$, with 
${\hat H}^\dagger={\hat H}$ and ${\hat{\mathit\Gamma}}^\dagger={\hat{\mathit\Gamma}}$, 
be a complex Hamiltonian with eigenstates $\{|\phi_n\rangle\}$ and eigenvalues $\{\kappa_n\}$: 
\begin{eqnarray}
{\hat K}|\phi_n\rangle = \kappa_n|\phi_n\rangle \quad {\rm and} \quad \langle\phi_n| 
{\hat K}^\dagger = {\bar\kappa}_n\langle\phi_n| . 
\label{eq:26}
\end{eqnarray} 
We shall assume for now that the eigenvalues $\{\kappa_n\}$ are not degenerate. In addition to 
the eigenstates of ${\hat K}$, it will be convenient to introduce eigenstates of the Hermitian adjoint 
matrix ${\hat K}^\dagger$: 
\begin{eqnarray}
{\hat K}^\dagger|\chi_n\rangle = \nu_n|\chi_n\rangle \quad {\rm and} \quad \langle\chi_n| 
{\hat K} = {\bar\nu}_n\langle\chi_n| . 
\label{eq:27}
\end{eqnarray} 
Here and in what follows, a `Hermitian adjoint' will be defined by the convention that 
${\hat K}^\dagger$ denotes the complex-conjugate transpose of ${\hat K}$. 
The reason for introducing the additional states $\{|\chi_n\rangle\}$ is because the eigenstates 
$\{|\phi_n\rangle\}$ of ${\hat K}$ are in general not orthogonal: 
\begin{eqnarray}
\langle\phi_m|\phi_n\rangle = 2\ri \frac{\langle\phi_m|{\hat{\mathit\Gamma}}|\phi_n\rangle}
{{\bar\kappa}_m-\kappa_n} = 2 \frac{\langle\phi_m|{\hat{H}}|\phi_n\rangle}
{{\bar\kappa}_m+\kappa_n}
\label{eq:28}
\end{eqnarray}
for $m\neq n$, which follows from the facts that $2\ri{\hat{\mathit\Gamma}} = {\hat K}^\dagger-
{\hat K}$ and that $2{\hat H}= {\hat K}^\dagger+{\hat K}$. An analogous result 
\begin{eqnarray}
\langle\chi_m|\chi_n\rangle = 2\ri \frac{\langle\chi_m|{\hat{\mathit\Gamma}}|\chi_n\rangle}
{\nu_n-{\bar\nu}_m} = 2 \frac{\langle\chi_m|{\hat{H}}|\chi_n\rangle}
{\nu_n+{\bar\nu}_m}
\label{eq:29}
\end{eqnarray}
holds for the eigenstates $\{|\chi_n\rangle\}$ of ${\hat K}^\dagger$. Of course, for a given 
${\hat K}$ some of its eigenstates can be orthogonal, but if ${\hat K}$ is not Hermitian, then a 
typical situation that arises is where not all the eigenstates are orthogonal. Hence conventional 
projection techniques so commonly used in many calculations of quantum mechanics, for 
example, in measurement theory or perturbation analysis, are ineffective when dealing with the 
eigenstates of a complex Hamiltonian \cite{more}. 

With the aid of the conjugate basis $\{|\chi_n\rangle\}$, let us first establish that the eigenstates 
$\{|\phi_n\rangle\}$ of ${\hat K}$, although not orthogonal, are nevertheless linearly independent. 
To show this, suppose the converse that $\{|\phi_n\rangle\}$ are linearly dependent. Then there 
exists a set of numbers $\{c_n\}$ such that $\sum_n|c_n|^2\neq0$, and that 
\begin{eqnarray}
\sum_n c_n |\phi_n\rangle = 0 . 
\label{eq:x40}
\end{eqnarray}
Transvecting this relation with $\langle\chi_m|$ from the left, we find, for each $m$, that $c_m 
\langle\chi_m|\phi_m\rangle=0$, where we have made use of the facts that 
\begin{eqnarray}
\langle\chi_n|\phi_m\rangle = \delta_{nm}\langle\chi_n|\phi_n\rangle 
\label{eq:30.0}
\end{eqnarray}
and that $\langle\chi_n|\phi_n\rangle\neq0$. To see that (\ref{eq:30.0}) holds, we note that by 
definitions (\ref{eq:26}) and (\ref{eq:27}) we have 
\begin{eqnarray}
\langle\chi_m|{\hat K}|\phi_n\rangle = {\bar\nu}_m\langle\chi_m|\phi_n\rangle = 
\kappa_n \langle\chi_m|\phi_n\rangle .
\label{eq:31}
\end{eqnarray} 
Hence $\langle\chi_m|\phi_n\rangle=0$ if $\kappa_n\neq{\bar\nu}_m$, and $\kappa_n=
{\bar\nu}_m$ if $\langle\chi_m|\phi_n\rangle\neq0$. Since $\langle\chi_m|\phi_n\rangle=0$ 
cannot hold for all $\{|\chi_m\rangle\}$, there has to be at least one $\nu_m$ such that 
$\kappa_n={\bar\nu}_m$. On the other hand, by assumption the eigenvalues are not degenerate, 
so there cannot be more than one $\nu_m$ for which $\kappa_n={\bar\nu}_m$. Without loss 
of generality we can label the states such that we have $\kappa_n={\bar\nu}_n$ for all $n$. 
It follows that $\langle\chi_m|\phi_n\rangle=0$ if $n\neq m$ but $\langle\chi_n|\phi_n\rangle
\neq0$, and this establishes (\ref{eq:30.0}). Now since $\langle\chi_m|\phi_m\rangle\neq0$ 
when ${\hat K}$ is nondegenerate, we must have $c_m=0$ for all $m$, contradicting the 
hypothesis. It follows that the nondegenerate eigenstates $\{|\phi_n\rangle\}$ of ${\hat K}$ 
are linearly independent, and thus span the Hilbert space ${\mathcal H}$, since the number 
of linearly independent basis elements agrees with the Hilbert-space dimensionality. In other 
words, $\{|\phi_n\rangle\}$ forms a \textit{complete} set of basis for ${\mathcal H}$. Additionally, 
they are \textit{minimal} in that exclusion of any one of the elements $|\phi_k\rangle$ from 
the set $\{|\phi_n\rangle\}$ spoils completeness. A set of basis elements that is both minimal 
and complete is called \textit{exact}. In finite dimensions, the exactness of $\{|\phi_n\rangle\}$ 
implies the exactness of $\{|\chi_n\rangle\}$, whereas in infinite dimensions this no longer is 
the case, as discussed below in \S\ref{sec:10}. 

Using the independence of the states $\{|\phi_n\rangle\}$ we can establish the relation: 
\begin{eqnarray}
\sum_n \frac{|\phi_n\rangle \langle\chi_n|}{\langle\chi_n|\phi_n\rangle} = {\mathds 1} ,
\label{eq:30}
\end{eqnarray}
which hold in finite dimensions away from degeneracies. To show this, 
we remark that if ${\hat F}$ has the property that $\langle\psi|{\hat F}|\psi\rangle=
\langle\psi|\psi\rangle$ holds true for an arbitrary vector $|\psi\rangle$, then it must be that 
${\hat F}={\mathds 1}$. Writing $|\psi\rangle=\sum_m c_m 
|\phi_m\rangle$ for some $\{c_m\}$ we have 
\begin{eqnarray}
\langle\psi| \left( \sum_n \frac{|\phi_n\rangle \langle\chi_n|}{\langle\chi_n|\phi_n\rangle} \right) 
|\psi\rangle = \sum_n \sum_m {\bar c}_m c_n \langle\phi_m|\phi_n\rangle = \langle\psi|\psi\rangle , 
\end{eqnarray}
and this establishes the claim. 

The operator ${\hat\Pi}_n$ defined by  (cf. \cite{des Cloizeaux})
\begin{eqnarray}
{\hat\Pi}_n = \frac{|\phi_n\rangle \langle\chi_n|}{\langle\chi_n|\phi_n\rangle} 
\end{eqnarray}
thus plays the role of a projection operator satisfying ${\hat\Pi}_n{\hat\Pi}_m=\delta_{nm}
{\hat\Pi}_n$. Although ${\hat\Pi}_n$ is not Hermitian, its eigenvalues 
are all zero, except one which is unity, for which the eigenstate is $|\phi_n\rangle$. Writing 
${\hat\Phi}_n=|\phi_n\rangle\langle\phi_n|/\langle\phi_n|\phi_n\rangle$ for the eigenstate 
projector we have 
\begin{eqnarray}
{\hat\Pi}_n {\hat\Phi}_n = {\hat\Phi}_n{\hat\Pi}_n = {\hat\Phi}_n . 
\end{eqnarray} 
It follows, in particular, that 
\begin{eqnarray}
({\mathds 1}-{\hat\Pi}_n)|\phi_n\rangle = ({\mathds 1}-{\hat\Pi}^\dagger_n)|\chi_n\rangle = 0. 
\end{eqnarray} 
While the complex Hamiltonian ${\hat K}$ does not admit the representation $\sum_n \kappa_n 
{\hat\Phi}_n$, due to the fact that ${\hat\Phi}_n {\hat\Phi}_m \neq \delta_{nm}{\hat\Phi}_m$, it 
nevertheless can be expressed in the form (cf. \cite{Cloizeaux}): 
\begin{eqnarray}
{\hat K} = \sum_n \kappa_n {\hat\Pi}_n . 
\label{eq:z30} 
\end{eqnarray}
It follows, furthermore, that if we write, for an arbitrary state $|\psi\rangle=\sum_m c_m 
|\phi_m\rangle$, 
\begin{eqnarray}
\psi_n^\chi = \frac{\langle\phi_n|\psi\rangle}{\sqrt{\langle\phi_n|\chi_n\rangle}} 
\quad {\rm and} \quad 
\psi_n^\phi =\frac{ \langle\chi_n|\psi\rangle}{\sqrt{\langle\chi_n|\phi_n\rangle}}, 
\end{eqnarray}
then we have 
\begin{eqnarray}
\langle\varphi|\psi\rangle = \sum_n {\bar\varphi}_n^\chi\psi_n^\phi.
\end{eqnarray}
A form of this result for real Hilbert-space vectors was obtained in \cite{Pell}. 


\section{Quantum probabilities}
\label{sec:3}

In the foregoing discussion we have not commented on the norm convention. In quantum 
theory, the norm of a state is closely related to probabilistic interpretations of measurement 
outcomes. Hence we wish 
to fix our norm convention so that it is consistent with probabilistic considerations of a 
quantum system when energy eigenstates are not orthogonal. Now in the literature on the use 
of biorthogonal basis for complex Hamiltonians, especially in quantum chemistry, the norm of 
the eigenvectors are often (but not always; cf. \cite{KKM,GRS} for a related discussion) 
assumed to take values larger than unity so as to ensure the following relation holds for all $n$:
\begin{eqnarray}
\langle\chi_n|\phi_n\rangle=1 .
\label{eq:zz33} 
\end{eqnarray}
Under this convention, eigenvectors will no longer be normalised. In particular, if we assume 
that all eigenstates have the same Hermitian norm so that $\langle\phi_n|\phi_n\rangle=\langle\phi_m
|\phi_m\rangle$ for all $n,m$, then we have $\langle\phi_n|\phi_n\rangle\geq1$. This might at first 
seem a little odd from the viewpoint of traditional Hermitian quantum mechanics, however, for a 
range of analysis that follow, it turns out that the convention $\langle\chi_n|\phi_n\rangle=1$ leads 
to considerable simplifications. 

To begin, we recall that in standard quantum mechanics, the `transition probability' between 
a pair of states $|\xi\rangle$ and $|\eta\rangle$ is given by the ratio of the form 
$\langle{\xi}|\eta\rangle\langle{\eta}|\xi\rangle/\langle{\xi}|\xi\rangle\langle{\eta}|\eta\rangle$. 
Under the convention 
$\langle\chi_n|\phi_n\rangle=1$, however, we cannot maintain a consistent probabilistic 
interpretation from this definition. For instance, if the state of the system is in an eigenstate 
$|\phi_n\rangle$ of a complex Hamiltonian ${\hat K}$, then on account of stationarity there 
cannot be a `transition' into another state $|\phi_m\rangle$, $m\neq n$, even though 
$\langle\phi_m|\phi_n\rangle\neq0$; whereas according to the conventional definition the 
transition probability between these states is nonzero. To reconcile these apparent 
contradictions we need the introduction of the so-called associated state that defines duality 
relations between elements of the Hilbert space ${\mathcal H}$ and its dual space 
${\mathcal H}^*$. 

For an arbitrary state $|\psi\rangle$, we define the \textit{associated state} $|{\tilde\psi}\rangle$ 
according to the following relations: 
\begin{eqnarray}
|\psi\rangle = \sum_n c_n |\phi_n\rangle \quad \Leftrightarrow \quad 
\langle{\tilde\psi}| = \sum_n {\bar c}_n \langle\chi_n| \quad \Rightarrow \quad 
|{\tilde\psi}\rangle = \sum_n c_n |\chi_n\rangle. 
\label{eq:q16}
\end{eqnarray}
We shall let (\ref{eq:q16}) determine the duality relation on the state space: 
$|\psi\rangle\in{\mathcal H} \Leftrightarrow |{\tilde\psi}\rangle \in {\mathcal H}^*$. Putting the 
matter differently, the state dual to $|\psi\rangle$ is given by $\langle{\tilde\psi}|$ of (\ref{eq:q16}); 
the state $|{\tilde\psi}\rangle$ associated to $|\psi\rangle$ is then given by the Hermitian 
conjugate of $\langle{\tilde\psi}|$. The quantum-mechanical inner product for a biorthogonal 
system is thus defined as follows: If $|\psi\rangle = \sum_n c_n |\phi_n\rangle$ and 
$|\varphi\rangle = \sum_n d_n |\phi_n\rangle$, then 
\begin{eqnarray}
\langle{\varphi},\psi\rangle \equiv 
\langle{\tilde\varphi}|\psi\rangle = \sum_{n,m} {\bar d}_n c_m \langle\chi_n|\phi_m \rangle 
= \sum_{n} {\bar d}_n c_n . \label{eq:HS} 
\end{eqnarray}
Since we demand the convention that $\langle\chi_n|\phi_n
\rangle=1$ for all $n$, we can assume that 
\begin{eqnarray}
\langle{\tilde\psi}|\psi\rangle=\sum_n{\bar c}_nc_n=1. 
\label{eq:q17}
\end{eqnarray}
It also follows that $p_n={\bar c}_nc_n$ defines the transition probability between 
$|\psi\rangle$ and $|\phi_n\rangle$: 
\begin{eqnarray}
p_n = \frac{\langle\chi_n|\psi\rangle \langle{\tilde\psi}|\phi_n\rangle}
{\langle{\tilde\psi}|\psi\rangle\langle\chi_n|\phi_n\rangle} ,
\label{eq:q18}
\end{eqnarray}
provided that the Hilbert space pairing is defined by the convention (\ref{eq:HS}). 
Here for definiteness we have expressed $p_n$ in a homogeneous form that is invariant under 
complex scale transformations of the states. The interpretation of the number $p_n$ is as follows: 
if a system is in a state characterised by the vector $|\psi\rangle$, and if a measurement is 
performed on the `complex observable' ${\hat K}$, then the probability that the measurement 
outcome taking the value $\kappa_n$ is given by $p_n$. 

More generally, the overlap distance $s$ between the two states $|\xi\rangle$ and $|\eta\rangle$ 
will be defined according to the prescription: 
\begin{eqnarray}
\cos^2 {\textstyle\frac{1}{2}} s = \frac{\langle{\tilde\xi}|\eta\rangle\langle{\tilde\eta}|\xi\rangle}
{\langle{\tilde\xi}|\xi\rangle\langle{\tilde\eta}|\eta\rangle} . 
\label{eq:q19}
\end{eqnarray}
A short exercise making use of the Cauchy-Schwarz inequality shows that the right side of 
(\ref{eq:q19}) is real, nonnegative, and lies between zero and one, thus qualifying the required 
probabilistic conditions. In particular, $s=0$ only if $|\xi\rangle=|\eta\rangle$; whereas $s=\pi$ 
only if $\sum_n{\bar c}_nd_n=0$ where $|\xi\rangle=\sum_nc_n|\phi_n\rangle$ and 
$|\eta\rangle=\sum_nd_n|\phi_n\rangle$. 

In quantum mechanics the notion of probability is closely related to that of distance. To see 
this, suppose that $|\eta\rangle=|\xi\rangle+|\rd\xi\rangle$ is a neighbouring state to $|\xi\rangle$. 
Then expanding (\ref{eq:q19}) and retaining terms of quadratic order, we obtain the following form 
of the line element, known as the Fubini-Study line element: 
\begin{eqnarray}
\rd s^2 = 4 \frac{\langle{\tilde\xi}|\xi\rangle\langle{\widetilde{\rd\xi}}|\rd\xi\rangle - 
\langle{\tilde\xi}|\rd\xi\rangle\langle{\widetilde{\rd\xi}}|\xi\rangle}
{\langle{\tilde\xi}|\xi\rangle^2} . 
\label{eq:52}
\end{eqnarray}
As an illustrative example, consider a two-dimensional Hilbert space spanned by a pair of states 
$(|\phi_1\rangle,|\phi_2\rangle)$. Then an arbitrary normalised---in the sense of (\ref{eq:q17})---state 
$|\xi\rangle$ can be expressed in the form 
\begin{eqnarray}
|\xi\rangle = \cos\half\theta|\phi_1\rangle + \sin\half\theta\re^{{\rm i}\varphi}|\phi_2\rangle. 
\label{eq:q21}
\end{eqnarray}
Evidently we have $\langle\xi|\xi\rangle\neq1$ but $\langle{\tilde\xi}|\xi\rangle=1$, on account 
of (\ref{eq:zz33}). Taking the differential of $|\xi\rangle$ and substituting the resulting expression 
in (\ref{eq:52}), making use of (\ref{eq:q16}), we deduce that the line element is given by
\begin{eqnarray}
\rd s^2 = \quat \left( \rd\theta^2 + \sin^2\theta \rd\varphi^2\right). 
\end{eqnarray}
It follows that the state space defined by the relation $\langle{\tilde\xi}|\xi\rangle=1$ is a 
two-sphere of radius one half---the Bloch sphere of complex Hamiltonian systems. We shall 
have more to say about this.


\section{Observables and states}
\label{sec:4}

We have shown in (\ref{eq:z30}) that a complex Hamiltonian ${\hat K}$ admits a spectral 
decomposition in terms of the complex projection operators $\{{\hat\Pi}_n\}$. Evidently, for a 
fixed biorthogonal basis $\{|\phi_n\rangle,|\chi_n\rangle\}$ there are uncountably many such 
(commuting family of) 
operators for which eigenvalues are entirely real, even though they are not Hermitian in the 
sense that ${\hat K}^\dagger$ does not agree with ${\hat K}$. In fact, 
the class of such `real' operators in this space is wider and contains those that do not 
commute with the Hamiltonian ${\hat K}$. 

Given a fixed biorthogonal basis $\{|\phi_n\rangle,|\chi_n\rangle\}$, a generic operator ${\hat F}$ 
can be expressed in the form 
\begin{eqnarray}
{\hat F} = \sum_{n,m} f_{nm} |\phi_n\rangle\langle\chi_m| . 
\label{eq:w41}
\end{eqnarray} 
Note that ${\hat F}$ can likewise be expressed in terms of the \textit{nonorthogonal} basis 
$\{|\phi_n\rangle\}$: 
\begin{eqnarray}
{\hat F} = \sum_{n,m} \varphi_{nm} |\phi_n\rangle\langle\phi_m| ,
\end{eqnarray} 
since the set $\{|\phi_n\rangle\}$ is complete. 
However, in this case the array $\{\varphi_{nm}\}$ cannot be viewed as a matrix, whereas the 
array $\{f_{nm}\}$ can, which shows the advantage of the use of biorthogonal basis. Thus, if 
${\hat G}$ is another operator with `matrix' elements $g_{nm}$ in the basis 
$\{|\phi_n\rangle,|\chi_n\rangle\}$, then the matrix element of the product ${\hat F}{\hat G}$ is 
just $\sum_lf_{nl}g_{lm}$. 

If ${\hat F}$ and ${\hat G}$ are nondegenerate Hermitian---in the usual sense---operators, the 
eigenstates of ${\hat F}$ can always be transformed unitarily into those of ${\hat G}$. For 
complex operators, however, this is no longer the case. Nevertheless, two operators ${\hat F}$ 
and ${\hat G}$ will be said to belong to the same class of observables if there is a unitary 
transformation between the basis of ${\hat F}$ and ${\hat G}$. 

The expectation value of a generic observable ${\hat F}$ in a pure state $|\psi\rangle$ is 
defined by the expression 
\begin{eqnarray}
\langle{\hat F}\rangle = \frac{\langle{\tilde\psi}|{\hat F}|\psi\rangle}{\langle{\tilde\psi}|\psi\rangle}. 
\label{eq:z54} 
\end{eqnarray} 
In particular, if the array $\{f_{nm}\}$ in (\ref{eq:w41}) is `biorthogonally Hermitian' in the sense that 
${\bar f}_{nm}=f_{mn}$, then $\langle{\hat F}\rangle$ defined by (\ref{eq:z54}) is real for all states 
$|\psi\rangle$, even though $\langle\psi|{\hat F}|\psi\rangle/\langle\psi|\psi\rangle$ is not real for most 
states. Thus, the notion of Hermiticity extends naturally to the biorthogonal setup, and we are able to 
speak about physical observables in the usual sense. This follows from the fact that although 
${\hat F}$ is not Hermitian in the sense that ${\hat F}^\dagger \neq {\hat F}$, its expectation value 
(\ref{eq:z54}) in an arbitrary state $|\psi\rangle$ is nevertheless real because the corresponding 
matrix $\{f_{nm}\}$ in the biorthogonal basis is Hermitian. If we let $|\psi\rangle=\sum_n c_n|\phi_n
\rangle$ and substitute this in (\ref{eq:z54}), making use of (\ref{eq:w41}), then we find 
\begin{eqnarray}
\langle{\hat F}\rangle =  \frac{\sum_{n,m} {\bar c}_n c_m f_{nm}}{\sum_n {\bar c}_n c_n} . 
\label{eq:z54-2} 
\end{eqnarray} 
In particular, if $\{|\phi_n\rangle\}$ are eigenstates of ${\hat F}$, then we can write $f_{nm}=f_n 
\delta_{nm}$, where $\{f_n\}$ are the eigenvalues of ${\hat F}$, hence 
\begin{eqnarray}
\langle{\hat F}\rangle =  \sum_{n} p_n f_n , 
\label{eq:z54-3} 
\end{eqnarray} 
which is consistent with our probabilistic interpretation of the biorthogonal system. 

The matrix interpretation here nevertheless requires further clarification. If a Hermitian `matrix' 
$f_{nm}$ is given without the information about the choice of basis, then there is no procedure 
to determine whether ${\hat F}$ is Hermitian; whereas for orthogonal bases, the data $f_{nm}$ 
is sufficient to determine whether ${\hat F}$ is Hermitian, even though the choice of the 
orthogonal basis remains arbitrary. To make this transparent, suppose that $\{|e_n\rangle\}$ 
is an orthonormal basis of ${\mathcal H}$ such that 
\begin{eqnarray}
|\phi_n\rangle = \sum_k u_n^k |e_k\rangle, \qquad |\chi_n\rangle = \sum_k v_n^k |e_k\rangle. 
\label{eq:q26} 
\end{eqnarray}
Then the matrix element of the observable ${\hat F}$ in this orthonormal basis is given by 
\begin{eqnarray}
{\hat F} = \sum_{n,m} \left( \sum_{k,l} f_{kl} u_k^n {\bar v}_l^m \right) |e_n\rangle\langle e_m| . 
\label{eq:w441}
\end{eqnarray} 
In this way we see more explicitly that while the reality of ${\hat F}$ merely requires Hermiticity 
of $\{f_{nm}\}$, the Hermiticity of ${\hat F}$ requires a more stringent condition that 
\begin{eqnarray}
\sum_{k,l}f_{kl}u_k^n{\bar v}_l^m=\sum_{k,l}{\bar f}_{kl}{\bar u}_k^m v_l^n. 
\label{eq:w442}
\end{eqnarray} 
In particular, if ${\hat F}$ is Hermitian so that ${\hat F}^\dagger={\hat F}$, then $\{|e_n\rangle\}$ 
can be chosen to be $|\phi_n\rangle$ 
so that $u_k^n=v_k^n=\delta_k^n$ and (\ref{eq:w442}) reduces to the familiar condition 
$f_{nm}={\bar f}_{mn}$; if ${\hat F}$ is symmetric, then the left side of (\ref{eq:w442}) is invariant 
under the interchange of indices $m\leftrightarrow n$, and we have $v_k^n={\bar u}_k^n$, i.e. 
components of $|\chi_n\rangle$ are complex conjugates of the components of $|\phi_n\rangle$. 
The expansion coefficients $\{u_k^n\}$ are unique up to unitary transformations. The 
linear independence of $\{|\phi_n\rangle\}$ implies that $\{u^k_n\}$ is invertible, and the 
orthonormality condition $\langle\chi_n|\phi_m\rangle=\delta_{nm}$ implies that the inverse of 
$\{u^k_n\}$ is given by $\{{\bar v}_n^k\}$. Phrased differently, if we write (\ref{eq:q26}) in the 
form $|\phi_n\rangle = {\hat u}|e_n\rangle$ and $|\chi_n\rangle = {\hat v}|e_n\rangle$, then we 
have ${\hat v}^\dagger{\hat u}={\mathds 1}$; if ${\hat F}$ is real (biorthogonally Hermitian), 
then 
\begin{eqnarray}
{\hat F}^\dagger = {\hat v}{\hat v}^\dagger\, {\hat F} \, {\hat u}{\hat u}^\dagger 
= ({\hat u}{\hat u}^\dagger)^{-1} {\hat F} \, ({\hat u}{\hat u}^\dagger) , 
\end{eqnarray}
where ${\hat u}{\hat u}^\dagger$ is an invertible positive Hermitian operator. 

As an elementary illustrative example, consider the complex $2\times2$ Hamiltonian 
${\hat K}={\hat\sigma}_x-\ri\gamma{\hat\sigma}_z$ with $\gamma^2<1$. A short calculation 
shows that the eigenstates of ${\hat K}$ and ${\hat K}^\dagger$, in the region $\gamma^2<1$ 
for which the eigenvalues $\pm\sqrt{1-\gamma^2}$ are real, are given by 
\begin{eqnarray}
|\phi_\pm\rangle = n_\pm \left( \begin{array}{c} 1 \\ \ri\gamma\pm\sqrt{1-\gamma^2} 
\end{array} \right) , \qquad 
|\chi_\pm\rangle = n_\mp \left( \begin{array}{c} 1 \\ -\ri\gamma\pm\sqrt{1-\gamma^2} 
\end{array} \right), 
\end{eqnarray}
where $n_\pm^2=(1\mp\ri\gamma/\sqrt{1-\gamma^2})/2$, and where we have written 
$|\phi_+\rangle$ for $|\phi_1\rangle$, and so on. An arbitrary observable for which the expectation 
value defined by (\ref{eq:z54}) is \textit{real} can be expressed, up to trace, as a linear combination 
of the deformed Pauli matrices 
\begin{eqnarray}
\hspace{-1.2cm} 
{\hat\sigma}_x^\gamma = \frac{1}{\sqrt{1-\gamma^2}}\left( \begin{array}{cc} -\ri\gamma & 1 \\ 
1 & \ri\gamma \end{array} \right), \quad 
{\hat\sigma}_y^\gamma = \left( \begin{array}{cc} 0 & -\ri \\ \ri & 0 \end{array} \right), \quad 
{\hat\sigma}_z^\gamma = \frac{1}{\sqrt{1-\gamma^2}}\left( \begin{array}{cc} 1 & \ri\gamma \\ 
\ri\gamma & -1 \end{array} \right). 
\end{eqnarray} 
These are obtained according to the prescriptions 
\begin{eqnarray}
\hspace{-2.0cm} 
{\hat\sigma}_x^\gamma=|\phi_1\rangle\langle\chi_2|+|\phi_2\rangle\langle\chi_1|, \quad 
{\hat\sigma}_y^\gamma=-\ri|\phi_1\rangle\langle\chi_2|+\ri|\phi_2\rangle\langle\chi_1|, \quad 
{\hat\sigma}_z^\gamma=|\phi_1\rangle\langle\chi_1|-|\phi_2\rangle\langle\chi_2|.
\label{eq:pauli}
\end{eqnarray}
It should be evident that the triplet $({\hat\sigma}_x^\gamma,{\hat\sigma}_y^\gamma, 
{\hat\sigma}_z^\gamma)$ fulfils the standard $\mathfrak{su}(2)$ commutation relations, and 
that in the Hermitian limit $\gamma\to0$ we recover the standard Pauli matrices. The 
expectation values, in the sense of (\ref{eq:z54}), of these Pauli matrices in a generic state 
(\ref{eq:q21}) are thus given by 
\begin{eqnarray}
\langle{\hat\sigma}_x^\gamma\rangle=\sin\theta\cos\varphi, \quad 
\langle{\hat\sigma}_y^\gamma\rangle=\sin\theta\sin\varphi, \quad 
\langle{\hat\sigma}_z^\gamma\rangle=\cos\theta.  
\label{eq:q32}
\end{eqnarray}
Note that the right-sides of these expectation values are independent of $\gamma$, on 
account of the $\gamma$-dependence of the eigenstates. Expectation values of Hermitian 
operators, such as the usual Pauli matrices, on the other hand, are in general not real since 
they do not represent physical observables in the biorthogonal system. 

It should be evident, incidentally, that in the case of a two-level system, the choice of the 
biorthogonal system $\{|\phi_{1,2}\rangle\}$ is uniquely determined by the overlap distance 
$\arccos|\langle\phi_1|\phi_2\rangle|$, up to unitarity. Physical observables constructed under 
the biorthogonal system $\{|\phi_{1,2}\rangle,|\chi_{1,2}\rangle\}$ therefore belong to the same 
class of observables as those constructed from another system $\{|\phi'_{1,2}\rangle,|\chi'_{1,2}
\rangle\}$, provided that $|\langle\phi_1|\phi_2\rangle|=|\langle\phi'_1|\phi'_2\rangle|$. 

We have spoken about pure states thus far, but the state of a physical system in quantum 
mechanics is, more generally, and perhaps more commonly, characterised by a mixed state 
density matrix:  
\begin{eqnarray}
{\hat\rho} = \sum_{n,m} \rho_{nm} |\phi_n\rangle\langle\chi_m| . 
\end{eqnarray} 
A density matrix ${\hat\rho}$ is thus not Hermitian in the usual sense so that ${\hat\rho}\neq 
{\hat\rho}^\dagger$, but it is `Hermitian' with respect to the choice of biorthogonal basis 
$\{|\phi_n\rangle,|\chi_n\rangle\}$ so that ${\bar\rho}_{nm}=\rho_{mn}$. 
The eigenvalues of ${\hat\rho}$ are nonnegative and add up to unity. The expectation value of 
a generic observable (\ref{eq:w41}) in the state ${\hat\rho}$ is thus defined by 
\begin{eqnarray}
\langle{\hat F}\rangle = \tr({\hat\rho}{\hat F}) = \sum_n \langle\chi_n|{\hat\rho}{\hat F}
|\phi_n\rangle = \sum_{n,m} \rho_{nm}f_{mn}. 
\end{eqnarray} 
It should be evident that a necessary and 
sufficient condition for the reality of $\langle{\hat F}\rangle$, for an arbitrary ${\hat\rho}$, is that 
${\bar f}_{nm}=f_{mn}$. 

A simple example of a density matrix arises if a quantum system described by a complex 
Hamiltonian ${\hat K}$ is immersed in a heat bath of inverse temperature $\beta$. In particular, 
if the eigenvalues $\{\kappa_n\}$ of ${\hat K}$ are all real, then after a passage of time the 
system will reach an equilibrium state 
\begin{eqnarray}
{\hat\rho} = \frac{\re^{-\beta{\hat K}}}{\tr(\re^{-\beta{\hat K}})} = 
\sum_n \re^{-\beta \kappa_n -\ln Z(\beta)} |\phi_n\rangle\langle\chi_n| ,
\label{eq:w32}
\end{eqnarray}
if we assume the postulate that an equilibrium state should maximise the von Neumann entropy 
$-\tr({\hat\rho}\ln{\hat\rho})$ subject to the constraint that the system must possess a 
definite energy expectation $\tr({\hat\rho}{\hat K})$. Here, 
$Z(\beta)=\tr(\re^{-\beta{\hat K}})$ denotes the partition function. The reality of all the 
eigenvalues of ${\hat K}$ is crucial for the existence of a canonical distribution (\ref{eq:w32}), 
owing to properties of the dynamics of the system, as described below in \S8.


\section{Measurement of spin-$\frac{1}{2}$ particle} 
\label{sec:5}

We now wish to turn to the discussion about the Bloch sphere introduced in \S\ref{sec:3} 
above, in the context of a spin-$\frac{1}{2}$ particle system in quantum mechanics. To this 
end we recall first with the general discussion that in standard nonrelativistic quantum 
mechanics, the wave function of a particle splits into two components, one associated with 
its spacial symmetry and 
the other associated with its internal symmetry (such as spin, isospin, colour, flavour, etc.). 
Since in the nonrelativistic context these spacial and internal symmetries are independent, 
if one is interested only in the internal symmetry of a particle, then it is a common practice 
to ignore the spacial degrees of freedom of the wave function (belonging to an 
infinite-dimensional Hilbert space) and focus attention on the internal symmetries (belonging 
to a finite-dimensional Hilbert space). It follows, in particular, that internal symmetries of a 
particle, \textit{a priori}, do not concern the spacial degrees of freedom. 

In spite of the independence of these symmetries, one commonly speaks, for instance, about 
the spin of an electron in a certain spacial direction. The reason why this is permissible has 
its origin in the mathematical structure of the state space of a spin-$\frac{1}{2}$ particle system: 
The space of states for this system is a two-sphere---in the quantum context this is often 
referred to as the Bloch sphere---which can be embedded in a three-dimensional Euclidean 
space ${\mathds R}^3$. The implication of this remarkable fact is that one may select an 
arbitrary point on the state space and declare this point to be, say, the `north pole'. In this 
manner, each spin degrees of freedom of 
a spin-$\frac{1}{2}$ particle is mapped, one-to-one, to a direction in three dimensions. This 
identification is sometimes referred to as the Pauli correspondence, and can be seen in 
different ways. For example, from (\ref{eq:q32}) one sees that the expectation value of a spin 
operator (which is one-half of the Pauli matrices) takes a value on a sphere of radius one-half 
in ${\mathds R}^3$ (see \cite{BG0,Mueller,Brukner} for further discussion on the relation 
between the spacial dimension of the space-time and the spin of quantum particles).

With this background of standard quantum mechanics in mind, let us now turn to a 
spin-$\frac{1}{2}$ particle characterised by a Hamiltonian ${\hat K}$ whose eigenstates are 
not orthogonal. The relevant mathematical machineries have already been introduced above, 
but let us introduce them here in a slightly different order: Rather than starting from a Hamiltonian 
${\hat K}$, let us start from the specification of the eigenstates. Specifically, suppose that a 
pair of distinct states $(|\phi_1\rangle,|\phi_2\rangle)$ is given in a two-dimensional Hilbert space 
${\mathcal H}$ such that $\langle\phi_1|\phi_2\rangle\neq0$. We then find the conjugate pair 
$(|\chi_1\rangle,|\chi_2\rangle)$ by solving the equations $\langle\chi_1|\phi_2\rangle=0$ and 
$\langle\chi_2|\phi_1\rangle=0$, satisfying the norm convention $\langle\chi_1|\phi_1\rangle=
\langle\chi_2|\phi_2\rangle=1$; solutions will be unique up to overall phases. We then identify 
the Hamiltonian according to 
\begin{eqnarray}
{\hat K}=\kappa_1|\phi_1\rangle\langle\chi_1|+\kappa_2
|\phi_2\rangle\langle\chi_2|, 
\end{eqnarray}
which, alternatively, can be expressed in the form ${\hat K}={\boldsymbol B}\cdot
{\hat{\boldsymbol\sigma}}$ for some choice of real vector ${\boldsymbol B}$, where 
${\boldsymbol\sigma}$ is the Pauli-matrix vector obtained by use of the biorthogonal basis, 
in accordance with (\ref{eq:pauli}). This Hamiltonian, although not Hermitian, nevertheless 
has the interpretation of representing the energy of a spin-$\frac{1}{2}$ particle system 
immersed in an external magnetic field ${\boldsymbol B}$ in ${\mathds R}^3$. 

This result follows from our probability assignment rule (\ref{eq:q19}). To see this, we recall that 
a generic state of the particle can be expressed in the form (\ref{eq:q21}). Now the spherical 
coordinates used in (\ref{eq:q21}) show that the two eigenstates $|\phi_1\rangle$ and 
$|\phi_2\rangle$ are \textit{antipodal} points on the Bloch sphere, even though they are not 
orthogonal in ${\mathcal H}$. We have explained that when an experimentalist performs a spin 
measurement, the direction of the measurement apparatus in ${\mathds R}^3$ is in one-to-one 
correspondence with the point on the Bloch sphere $S^2$, not so much with the direction in 
Hilbert space ${\mathcal H}$ as such, in the chain of abstraction ${\mathds R}^3\to S^2 \to 
{\mathcal H}$. To put the matter differently, the data obtained from the Stern-Gerlach experiment 
(see \cite{FH} for a curious historical account of the experiment) does not provide information 
concerning whether the `spin-up' state and `spin-down' state correspond to orthogonal vectors 
in ${\mathcal H}$; it merely tells us that they correspond to antipodal points on $S^2$, whereas 
going from $S^2$ to ${\mathcal H}$ requires further milages requiring more information than 
mere experimental data.  

For sure the use of orthogonal bases---hence the use of Hermitian operators---simplifies the 
algebra, but apart from this `convenience' argument, there is no need to require orthogonality 
in ${\mathcal H}$; all that is needed is the completeness. We are therefore led to the following 
conclusion: 

\begin{prop} 
In finite dimensions, the interrelation, i.e. the overlap distances, of the eigenstates of 
nondegenerate observables with real eigenvalues in Hilbert space cannot be determined 
from experimental data. 
\label{prop:1}
\end{prop} 

In other words, any operator possessing the relevant eigenvalue structure is a legitimate 
candidate for a physical observable. Hence Hermitian operators have no privileged status, 
apart from their ability in making calculations simpler. 
This conclusion, however, is not necessarily true in infinite dimensions; likewise in finite 
dimensions, one can identify differences between Hermitian and non-Hermitian observables 
if at least one of the eigenvalues is complex, or if there are degeneracies of eigenstates. We 
shall have more to say about these points.


\section{Spin particles and combined systems} 
\label{sec:6}

Particles with higher spin numbers can be formulated analogously. Of course, one might ask, 
even in the case of standard quantum mechanics with Hermitian observables, in which way 
spin measurements in ${\mathds R}^3$ can be related to points on the state space since the 
dimensionality of the state space for higher spin systems is larger than three and hence it 
cannot be embedded in ${\mathds R}^3$. The way to realise the Pauli correspondence for 
higher spin systems is to note the fact that in the state space for each spin, there is a family 
of privileged quantum states, sometimes called the $\mathfrak{su}(2)$ coherent states, that 
fully embody information concerning directional data in ${\mathds R}^3$ (see \cite{BH,BG1} for 
a detailed discussion), and that the coherent state subspace is always a two sphere $S^2$ that 
can be embedded in ${\mathds R}^3$. It is via this device that the idea of the Pauli 
correspondence for spin-$\frac{1}{2}$ particle can be extended to arbitrary spin particles. 
To put the matter differently, for higher spins there is a natural embedding of the directional 
data of ${\mathds R}^3$ in the state space of the system. 

It should be evident from the discussion of the preceding section that a similar line of reasoning 
is applicable to biorthogonal quantum systems. As an example, consider a spin-$\frac{1}{2}$ 
state vector $|\psi\rangle=c_1|\phi_1\rangle+c_2|\phi_2\rangle$ in ${\mathcal H}^2$, normalised 
as usual according to $\langle {\tilde\psi}|\psi\rangle=1$. We embed this state in ${\mathcal H}^3$ 
by consideration of the product state:
\begin{eqnarray}
|\psi,\psi\rangle = c_1^2|\phi_1,\phi_1\rangle + \sqrt{2}c_1c_2 \left( \frac{|\phi_1,\phi_2\rangle 
+ |\phi_2,\phi_1\rangle}{\sqrt{2}} \right) + c_2^2|\phi_2,\phi_2\rangle . 
\end{eqnarray}
This coherent state in ${\mathcal H}^3$ is then identified as the spin-$1$ state in some direction 
of ${\mathds R}^3$, which becomes more apparent if we choose the parameterisation 
$c_1=\cos\frac{1}{2}\theta$ and $c_2=\sin\frac{1}{2}\theta\,\re^{{\rm i}\varphi}$. Clearly 
$|\psi,\psi\rangle$ is normalised in the sense of (\ref{eq:q17}) since $|c_1|^2+|c_2|^2=1$. 
If we call $\theta
=0$ the positive $z$-direction in ${\mathds R}^3$, then the triplet of states 
\[
\left( |\phi_1,\phi_1\rangle, \quad \frac{|\phi_1,\phi_2\rangle + |\phi_2,\phi_1\rangle}{\sqrt{2}} , 
\quad  |\phi_2,\phi_2\rangle \right)   
\]
corresponds to the three spin-$1$ eigenstates of $S_z$: 
\[
\Big( |S_z=+1\rangle, \quad |S_z=0\rangle, \quad  |S_z=-1\rangle \Big)   . 
\]
An arbitrary state of the spin-$1$ particle is therefore expressed as a liner combination of these 
basis states. 

This line of construction extends to all higher spin particles. Thus, for example, for a 
spin-$\frac{3}{2}$ system we form the coherent state 
\begin{eqnarray}
\hspace{-1.5cm} 
|\psi,\psi,\psi\rangle &=& c_1^3|\phi_1,\phi_1,\phi_1\rangle + \sqrt{3}c_1^2c_2 \left( 
\frac{|\phi_1,\phi_1,\phi_2\rangle + |\phi_1,\phi_2,\phi_1\rangle+ |\phi_2,\phi_1,\phi_1\rangle}
{\sqrt{3}} \right) \nonumber \\ & & + \sqrt{3}c_1c_2^2 \left( 
\frac{|\phi_1,\phi_2,\phi_2\rangle + |\phi_2,\phi_1,\phi_2\rangle+ |\phi_2,\phi_2,\phi_1\rangle}
{\sqrt{3}} \right) + c_2^3|\phi_2,\phi_2,\phi_2\rangle 
\end{eqnarray}
in ${\mathcal H}^4$ associated with $|\psi\rangle\in{\mathcal H}^2$, 
and identify the four states appearing here as the four eigenstates of the spin operator, and 
so on. 

The formulation presented here is somewhat unduly rigid in that if we define a $2\times2$ 
Hermitian matrix $\eta_{ij}=\langle\phi_i|\phi_j\rangle$, then the Hermitian transition 
amplitudes---as opposed to the physical transition amplitudes specified by 
(\ref{eq:q19})---between the spin eigenstates for all higher spins are entirely specified by the 
$2\times2$ matrix $\{\eta_{ij}\}$. In other words, the 
biorthogonal system for all higher spin systems are fixed once we fix that of the underlying 
spin-$\frac{1}{2}$ system. This rigidity, however, can in fact be relaxed, on account of 
Proposition~\ref{prop:1}, which shows that Hilbert space vectors play less prominent role 
than one might have thought. In particular, in biorthogonal quantum mechanics a coherent 
state can be constructed from incoherent Hilbert space vectors that are nevertheless 
projectively coherent. Thus, if 
$|\psi\rangle=c_1|\phi_1\rangle+c_2|\phi_2\rangle$ is given as before and if we define 
$|\psi'\rangle=c_1|\phi_1'\rangle+c_2|\phi_2'\rangle$, where $\langle\phi_i|\phi_j\rangle\neq 
\langle\phi_i'|\phi_j'\rangle$ so that $|\psi\rangle$ and $|\psi'\rangle$ are inequivalent Hilbert 
space vectors, then we can still form an admissible coherent state according to 
$|\psi,\psi'\rangle$. This follows on account of the fact that 
$\langle\chi_k|\psi\rangle=\langle\chi'_k|\psi'\rangle$, $k=1,2$, hence $|\psi\rangle$ and 
$|\psi'\rangle$ are projectively equivalent under our scheme. In this way we see that the 
biorthogonal basis for each spin particle can be chosen arbitrarily, without constraints. 

The observation made in the previous paragraph also shows that in biorthogonal quantum 
theory an arbitrary pair of systems can be combined without constraints. This, in turn, 
clarifies one of the outstanding issues of combined systems in PT-symmetric quantum 
mechanics, which we shall discuss later. For now it suffices to note that if one system 
represented by a Hilbert space ${\mathcal H}$ and another system represented by a Hilbert 
space ${\mathcal H}'$ are combined, then the state vector of the combined system is an 
element of the tensor product space ${\mathcal H}\otimes{\mathcal H}'$, just as in Hermitian 
quantum mechanics. Thus, for example, if $|\psi\rangle=c_1|\phi_1\rangle+c_2|\phi_2\rangle$ 
is the state of one spin-$\frac{1}{2}$ particle, and $|\psi'\rangle=c_1'|\phi_1'\rangle+c_2'
|\phi_2'\rangle$ is the state of another such particle, then a disentangled product state in 
${\mathcal H}\otimes{\mathcal H}'$ takes the form  
\begin{eqnarray}
|\psi,\psi'\rangle=c_1c_1'|\phi_1,\phi_1'\rangle + c_1c_2' |\phi_1,\phi_2'\rangle + c_2 c_1' 
|\phi_2,\phi_1'\rangle + c_2c_2'|\phi_2,\phi_2'\rangle, 
\end{eqnarray}
whereas a typical entangled state, such as the spin-$0$ singlet state, will be given by 
\begin{eqnarray}
|S=0,S_z=0\rangle=\frac{1}{\sqrt{2}}\, \Big( |\phi_1,\phi_2'\rangle - |\phi_2,\phi_1'\rangle \Big).
\label{eq:q41}
\end{eqnarray}
This might appear paradoxical at first, since the singlet state has to be antisymmetric, which 
is not immediately apparent from the right side of (\ref{eq:q41}). Indeed, $|\phi_n\rangle$ and 
$|\phi_n'\rangle$ represent distinct states in ${\mathcal H}$, however, they are projectively 
equivalent, which in turn makes (\ref{eq:q41}) antisymmetric in the projective Hilbert space. 

For a combined system, the interaction Hamiltonian can also be represented in a manner 
analogous to that in standard quantum mechanics. Thus, in the case of a pair of biorthogonal 
systems represented by a pair of Hamiltonians ${\hat K}={\hat\sigma}_x-\ri\gamma 
{\hat\sigma}_z$ and ${\hat K}'={\hat\sigma}_x-\ri\gamma'{\hat\sigma}_z$ with $\gamma^2, 
\gamma'^2<1$, the quantum Ising spin-spin interaction Hamiltonian can be expressed in the 
form 
\begin{eqnarray}
{\hat\sigma}_z^\gamma \otimes {\hat\sigma}_z^{\gamma'} = 
\frac{1}{\sqrt{(1-\gamma^2)(1-\gamma^{\prime2})}} \left( \begin{array}{cccc} 
1 & \ri\gamma' & \ri\gamma & -\gamma\gamma' \\ 
\ri\gamma' & -1 & -\gamma\gamma' & -\ri\gamma \\ 
\ri\gamma & -\gamma\gamma' & -1 & -\ri\gamma' \\ 
-\gamma\gamma' & -\ri\gamma & -\ri\gamma' & 1 
\end{array} \right), 
\label{eq:q42}
\end{eqnarray} 
whose eigenvalues are, of course, given by $(1,-1,1,-1)$, independent of $\gamma,\gamma'$.


\section{Perturbation analysis}
\label{sec:7}

We shall now turn to the perturbation analysis involving complex Hamiltonians, in the range 
where there are no degeneracies so that the Rayleigh-Schr\"odinger series is applicable. There 
is a substantial literature on perturbation theory involving complex Hamiltonians, even in the 
vicinities of degeneracies where not only eigenvalues but also eigenstates can be degenerate 
(see, for example, \cite{more,SW,kato,RS,FS}). As such, we have little new to add in this 
section, except perhaps the discussion on the nature of the operator that generates the 
perturbation, which turns out not to be unitary. 

Let ${\hat K}$ be a complex Hamiltonian with distinct eigenvalues $\{\kappa_n\}$ and biorthonormal 
eigenstates $(\{|\phi_n\rangle\},\{|\chi_n\rangle\})$ that are known. Suppose that we perturb the 
Hamiltonian slightly according to 
\begin{eqnarray}
{\hat K} \to {\hat K}_\epsilon = {\hat K}+\epsilon {\hat K}' , 
\end{eqnarray}
where $\epsilon\ll1$ is the perturbation parameter, and ${\hat K}'$ represents perturbation 
energy, which may or may not be Hermitian. Under the assumption that there are no degeneracies, 
the eigenstates $\{|\psi_n\rangle\}$ and the eigenvalues $\{\mu_n\}$ of the perturbed Hamiltonian 
${\hat K}_\epsilon$ can be expanded in a power series
\begin{eqnarray}
|\psi_n\rangle = |\phi_n\rangle + \epsilon |\psi_n^{(1)}\rangle + \epsilon^2 |\psi_n^{(2)}\rangle + 
\cdots, \quad \mu_n = \kappa_n + \epsilon \mu_n^{(1)} + \epsilon^2 \mu_n^{(2)} + \cdots .
\label{eq:q43} 
\end{eqnarray}
As for the normalisation of the perturbed eigenstates, we shall assume that 
\begin{eqnarray}
\langle\chi_n|\psi_n\rangle=1 . 
\end{eqnarray}
Since $\langle\chi_n|\phi_n\rangle=1$, it follows that under this normalisation convention we require 
\begin{eqnarray}
\langle\chi_n|\psi_n^{(1)}\rangle=\langle\chi_n|\psi_n^{(2)}\rangle=\cdots=0 . 
\label{eq:q45}  
\end{eqnarray}
It also means that $\langle{\tilde\psi_n}|\psi_n\rangle\neq1$, but the deviation from unity is negligible 
for $\epsilon\ll1$. 

If we substitute the series expansion (\ref{eq:q43}) in the eigenvalue equation 
\begin{eqnarray}
{\hat K}_\epsilon |\psi_n\rangle=\mu_n|\psi_n\rangle 
\label{eq:q46}  
\end{eqnarray}
and equate terms of different orders in $\epsilon$, then we 
obtain 
\begin{eqnarray}
(\kappa_n-{\hat K})|\phi_n\rangle = 0, \qquad 
(\kappa_n-{\hat K})|\psi_n^{(1)}\rangle + \mu_n^{(1)}|\phi_n\rangle={\hat K}'|\phi_n\rangle,
\label{eq:q47}  
\end{eqnarray}
and so on. Transvecting $\langle\chi_m|$ from the left on the second equation of (\ref{eq:q47}) 
we obtain 
\begin{eqnarray}
(\kappa_n-\kappa_m)\langle\chi_m|\psi_n^{(1)}\rangle + \mu_n^{(1)} \delta_{nm} = 
\langle\chi_m|{\hat K}'|\phi_n\rangle . 
\end{eqnarray}
Thus, for $n=m$ we obtain the first-order perturbation correction to the eigenvalue: 
\begin{eqnarray}
\mu_n^{(1)} = \langle\chi_n|{\hat K}'|\phi_n\rangle . 
\end{eqnarray}
On the other hand, for $n\neq m$ we obtain 
\begin{eqnarray}
\langle\chi_m|\psi_n^{(1)}\rangle = \frac{1}{\kappa_n-\kappa_m}\,
\langle\chi_m|{\hat K}'|\phi_n\rangle , 
\end{eqnarray}
and on account of the completeness condition we thus find 
\begin{eqnarray}
|\psi_n^{(1)}\rangle = \sum_{m} |\phi_m\rangle \langle\chi_m|\psi_n^{(1)}\rangle = 
\sum_{m\neq n} |\phi_m\rangle \langle\chi_m|\psi_n^{(1)}\rangle = \sum_{m\neq n} 
\frac{\langle\chi_m|{\hat K}'|\phi_n\rangle}{\kappa_n-\kappa_m}\, |\phi_m\rangle, 
\label{eq:q51} 
\end{eqnarray}
where we have made use of the orthogonality relations (\ref{eq:q45}). The results of \cite{SW} 
reproduced here for the first-order perturbation expansion lends itself naturally with the 
analysis of geometric phases for complex Hamiltonians \cite{GW,MKS,mostafa08,CZ}. 

It should be evident that higher-order perturbation corrections can be obtained in a manner 
analogous to the standard perturbation theory in Hermitian quantum mechanics, except the 
obvious modifications involving the biorthogonal basis elements. An important difference between 
(\ref{eq:q51}) and the conventional result, however, is that instead of the orthogonality condition 
$\langle\phi_n|\psi_n^{(1)}\rangle=0$, here we have $\langle\chi_n|\psi_n^{(1)}\rangle=0$. Now 
suppose that we regard ${\hat K}_\epsilon$ for $|\epsilon|\ll1$ as a one-parameter family of 
Hamiltonians connected 
to, and in the vicinity of, ${\hat K}$. Then the eigenstates $|\psi_n\rangle$ for a small range of 
$\epsilon$ constitutes a segment of a path in ${\mathcal H}$. If ${\hat K}$ is Hermitian, then a 
small displacement along the path is unitary, and leaves the norm of the eigenstate invariant. In the 
present context, the displacement is generated by the operator 
\begin{eqnarray}
{\hat V} = \sum_n |\psi_n(\epsilon)\rangle\langle\chi_n| ,
\label{eq:q52}
\end{eqnarray}
where we have written $|\psi_n(\epsilon)\rangle$ to make the $\epsilon$ dependence more 
explicit. In other words, we have ${\hat V}|\phi_n\rangle=|\psi_n\rangle$. Evidently, ${\hat V}$ is 
not unitary, and hence its generator $\ri (\partial_\epsilon {\hat V}){\hat V}^{-1}$ is not Hermitian. 
In particular, perturbation of an eigenstate $|\phi_n\rangle$ of a complex Hamiltonian ${\hat K}$ 
does not leave the Dirac norm $\langle\phi_n|\phi_n\rangle$ of the state invariant, but instead 
leaves invariant the biorthogonal norm $\langle\chi_n|\phi_n\rangle$ of the state, and this in turn 
gives another support for the use of (\ref{eq:q19}) as determining the physical probability rules 
involving complex Hamiltonians. 

We remark, incidentally, that in the case of a Hermitian operator, a theorem of Rellich implies that 
the eigenstates and eigenvalues can be expanded in a Taylor series of the form (\ref{eq:q43}). 
However, for a general complex operator, the foregoing perturbation expansion breaks down in 
the vicinities of degeneracies where not only the eigenvalues but also the corresponding 
eigenstates coalesce. Such degeneracies are often referred to as `exceptional points' in the 
literature (see \cite{Heiss} and references cited therein), with nontrivial observational 
consequences \cite{Richter,Richter2}. Although the formal series expansion (\ref{eq:q43}) breaks 
down in the neighbourhood of an exceptional point, a perturbative analysis can nevertheless be 
pursued by employing the Newton-Puiseux series (\cite{RS}, Theorem XII.2, \cite{SM}), as 
employed, e.g., in \cite{GRS,GGKN,DG}. 

\section{Dynamics} 
\label{sec:8}

Thus far we have been considering static aspects of the eigenvalues and eigenstates of a complex 
Hamiltonian ${\hat K}$. We shall now turn to the analysis of the time evolution of a quantum state 
generated by such ${\hat K}$, in the context of time-independent Hamiltonians. Specifically, we 
consider properties of the evolution operator 
\begin{eqnarray}
{\hat U} = \re^{-{\rm i}{\hat K}t}, 
\end{eqnarray}
in units $\hbar=1$. Evidently, ${\hat U}$ is not unitary: ${\hat U}^\dagger{\hat U}\neq{\mathds 1}$. 
However, as we shall show, if the eigenvalues of ${\hat K}$ are real, then ${\hat U}$ in effect is 
unitary in the sense of biorthogonal quantum mechanics so that the norms of states and transition 
probabilities are preserved under the time evolution. 

It should be apparent that the solution to the dynamical equation 
\begin{eqnarray}
\ri \partial_t |\psi\rangle = {\hat K}|\psi\rangle,
\label{eq:q54} 
\end{eqnarray}
with initial condition $|\psi_0\rangle=\sum_nc_n|\phi_n\rangle$, is given by 
\begin{eqnarray}
|\psi_t\rangle = \sum_n c_n \re^{-{\rm i}\kappa_n t} |\phi_n\rangle.
\end{eqnarray}
According to our conjugation rule (\ref{eq:q16}) we thus have 
\begin{eqnarray}
\langle{\tilde\psi_t}| = \sum_n {\bar c}_n \re^{{\rm i}{\bar\kappa}_n t} \langle\chi_n| 
\quad \Rightarrow \quad 
|{\tilde\psi_t}\rangle = \sum_n c_n \re^{-{\rm i}{\kappa}_n t} |\chi_n\rangle. 
\label{eq:q56} 
\end{eqnarray}
The time-dependent biorthogonal norm of the state therefore is given by 
\begin{eqnarray}
\langle{\tilde\psi_t}|\psi_t\rangle = \sum_n {\bar c}_n c_n 
\re^{-{\rm i}(\kappa_n-{\bar\kappa}_n)t} . 
\end{eqnarray}
We thus see that if the eigenvalues of ${\hat K}$ are real so that ${\bar\kappa}_n=\kappa_n$, then 
for all time $t>0$ we have $\langle{\tilde\psi_t}|\psi_t\rangle=\langle{\tilde\psi_0}|\psi_0\rangle$.  
More generally, if ${\bar\kappa}_n=\kappa_n$, and if $|\varphi_t\rangle$ is also a solution to the 
Schr\"odinger equation (\ref{eq:q54}) with a different initial condition, then we have 
\begin{eqnarray}
\langle{\tilde\varphi_t}|\psi_t\rangle = \langle{\tilde\varphi_0}|\psi_0\rangle 
\end{eqnarray}
for all $t>0$. It follows that: 

\begin{prop} 
If the eigenvalues of ${\hat K}$ are real, then the time evolution operator $\re^{-{\rm i}{\hat K}t}$ 
is unitary with respect to the biorthogonal basis of ${\hat K}$, preserving the biorthogonal norms 
of the states and the transition probabilities between states. 
\end{prop}

Additionally, if the eigenvalues $\{\kappa_n\}$ are real, then $|{\tilde\psi_t}\rangle$ can be seen to 
satisfy the Schr\"odinger equation $\ri \partial_t |{\tilde\psi}\rangle = {\hat K}^\dagger
|{\tilde\psi}\rangle$ with the Hermitian-conjugated Hamiltonian ${\hat K}^\dagger$. This, however, 
is not generally true if at least one of the eigenvalues of ${\hat K}$ is not real: $\ri \partial_t 
|{\tilde\psi}\rangle \neq {\hat K}^\dagger|{\tilde\psi}\rangle$ in general, which can be seen from 
(\ref{eq:q56}). 

When one or more of the eigenvalues are imaginary or complex, then we have different 
characteristics for the dynamical behaviour of a quantum state. Let us write 
\begin{eqnarray}
\kappa_n = E_n -\ri \gamma_n 
\end{eqnarray} 
for the eigenvalues, where $\{E_n\}$ and $\{\gamma_n\}$ are real. Then we have 
\begin{eqnarray}
\langle{\tilde\psi_t}|\psi_t\rangle = \sum_n {\bar c}_n c_n \re^{-2\gamma_n t} = 
{\bar c}_{n_*} c_{n_*} \re^{-2\gamma_{n_*} t} \left( 1 + \sum_{n\neq n_*} 
\frac{{\bar c}_n c_n}{{\bar c}_{n_*} c_{n_*}}\, \re^{-2(\gamma_n-\gamma_{n_*})t} \right), 
\label{eq:q60} 
\end{eqnarray}
where $n_*$ is the value of $n$ such that $\gamma_n$ has the smallest value (amongst the terms 
in the expansion for which $c_n\neq0$). In most physical setups, $\gamma_n\geq0$, and an 
arbitrary initial state will decay into the state with the smallest $\gamma_n$ value, while at the same 
time the overall norm decays. This situation describes the behaviour of a particle trapped in a finite 
potential well; the norm $\langle{\tilde\psi_t}|\psi_t\rangle$ then describes the probability that the 
particle has not tunnelled out of the well. Note that if we let $c_n=\delta_{nk}$ in (\ref{eq:q60}) for 
some $k$, then we see that an eigenstate $|\phi_k\rangle$ of ${\hat K}$ for which $\gamma_k\neq0$ 
is not a stationary state, i.e. if $|\psi_0 \rangle=|\phi_k\rangle$, then $\langle{\tilde\psi_t}|\psi_t
\rangle =\re^{-2\gamma_k t}$. 

The fact that when the eigenvalues are complex the state with the slowest decay will in time 
dominate is of course well known in the context of systems with decays, but it is worth remarking 
that as a consequence when such a system is immersed in a heat bath, it cannot result in an 
equilibrium configuration characterised by the thermal state (\ref{eq:w32}). 

With the notion of dynamics we are in a position to discuss time reversibility. In standard 
quantum mechanics there is no ``one-size fits all'' notion of the action of time reversal operator 
(cf. \cite{MTR}). 
Furthermore, the action of time reversal operator is sometimes viewed as an antilinear map 
(a quadratic form) from the Hilbert space to its dual space: ${\mathcal H}\to{\mathcal H}^*$; and 
sometimes as an antilinear map (an operator) from Hilbert space to itself: ${\mathcal H} \to 
{\mathcal H}$. Here we shall consider the latter convention, in line with \cite{wigner}. With the aid of 
a time-reversal operator ${\mathcal T}$ we can establish, for example, the following \textit{geometric} 
identity 
\begin{eqnarray}
\langle\phi_m|\phi_n\rangle = \langle\chi_n|\chi_m\rangle
\label{eq:q62}
\end{eqnarray}
using the \textit{physical} argument analogous to that presented in \cite{SW}. Suppose that we 
let a state evolve in time under the Hamiltonian ${\hat K}$. From (\ref{eq:q60}) the decay rate of 
$|\phi_n\rangle$ is given by $2\gamma_n$, whereas from (\ref{eq:28}) we have 
\begin{eqnarray}
\gamma_n =  \frac{\langle\phi_n|{\hat{\mathit\Gamma}}|\phi_n\rangle}{\langle\phi_n|\phi_n\rangle}. 
\end{eqnarray}
In other words, the decay rate of $|\phi_n\rangle$ is determined by $\hat{\mathit{\Gamma}}$ (even 
though $\gamma_n$ is \textit{not} the physical expectation of $\hat{\mathit{\Gamma}}$ in the state 
$|\phi_n\rangle$). Since the time-reversed dynamics must be such that the state $|\phi_n\rangle$ 
grows at the same rate $2\gamma_n$, it follows that the time reversal 
operator ${\mathcal T}$ reverses the sign of $\ri\hat{\mathit{\Gamma}}$ but leaves $\hat{H}$ and 
$\hat{\mathit{\Gamma}}$ invariant: ${\mathcal T}{\hat K}{\mathcal T}^{-1}={\hat K}^\dagger$. In other 
words, ${\hat K}^\dagger{\mathcal T}={\mathcal T}{\hat K}$. Hence if we define 
\begin{eqnarray}
|\chi_n\rangle = {\mathcal T}|\phi_n\rangle, 
\label{eq:z39}
\end{eqnarray}
we find that $|\chi_n\rangle$ is the eigenstate of ${\hat K}^\dagger$ with eigenvalue 
${\bar\kappa}_n$. The identity (\ref{eq:q62}) then follows at once.

\section{Relation to PT symmetry} 
\label{sec:9}

As we have indicated earlier, interests in the study of classical and quantum systems described by 
complex, non-Hermitian Hamiltonians have increased significantly since the realisation by Bender 
and Boettcher \cite{BB} that a wide class of complex Hamiltonians possessing certain anti-linear 
symmetries can have entirely real eigenvalues. Specifically, the anti-linear symmetry considered in 
this context is that associated with the space-time inversion, i.e. parity-time (PT) reversal operation. 
Since the literature in the area of PT-symmetric quantum theory is substantial, and since some of 
the ideas relating to biorthogonal quantum mechanics outlined here have been identified directly or 
indirectly in the investigation of PT symmetry \cite{Curtright}, it will be useful to draw a special 
attention to the subject here. 

We begin this discussion by recalling that, if we write ${\hat\varg}=({\hat u}{\hat u}^\dagger)^{-1}$, 
then on account of (\ref{eq:q26}) we have 
\begin{eqnarray}
\langle e_n|e_n\rangle = \langle\phi_n| {\hat\varg} |\phi_n\rangle=1
\end{eqnarray}
for all $n$, where ${\hat\varg}$ by construction is an invertible positive Hermitian operator, which is 
unique and can be determined from the eigenstates \cite{mostafa1}: 
\begin{eqnarray}
{\hat\varg}^{-1} = \sum_n |\phi_n\rangle\langle\phi_n| . 
\end{eqnarray}
In addition, observe, for all $n$, that 
\begin{eqnarray}
\langle\phi_n| {\hat\varg}^2|\phi_n\rangle =  \langle e_n| ({\hat u}^{-1})^\dagger {\hat u}^{-1} 
({\hat u}^{-1})^\dagger {\hat u}^{-1} |e_n\rangle 
= \langle e_n| {\hat u}^{-1}({\hat u}^{-1})^\dagger|e_n\rangle 
= \langle\chi_n|\chi_n\rangle , 
\end{eqnarray}
but (\ref{eq:q62}) shows that $\langle\chi_n|\chi_n\rangle=\langle\phi_n|\phi_n\rangle$, so that 
${\hat\varg}$ is an involution: 
\begin{eqnarray}
{\hat\varg}^2={\mathds 1}. 
\end{eqnarray}
Perceived from the viewpoint of Hermitian inner-product space, 
therefore, the operator ${\hat\varg}$ plays the role of a `metric' for the Hilbert space. For example, 
the expectation value of a physical observable ${\hat F}$ can be written in the form
\begin{eqnarray}
\frac{\langle{\tilde\psi}| {\hat F}|\psi\rangle}{\langle{\tilde\psi}|\psi\rangle} = 
\frac{\langle{\psi}|{\hat\varg}{\hat F}|\psi\rangle}{\langle\psi| {\hat\varg}|\psi\rangle} 
\end{eqnarray}
that involves the metric operator under the Hermitian pairing. 

We see therefore that biorthogonal quantum mechanics can alternatively be viewed as 
`conventional' Hermitian quantum mechanics, but where Hilbert space is endowed with a nontrivial 
metric operator ${\hat\varg}$. As remarked in \S\ref{sec:1}, there are indeed proposals to equip 
Hilbert space with a nontrivial metric \cite{SGH,GHS}. The statement of Proposition~\ref{prop:1}, 
however, shows that for a physical system modelled on a finite-dimensional Hilbert space with a 
family of observables having real eigenvalues, there are no observable consequences associated 
with the choice of the metric ${\hat\varg}$. Since any choice of ${\hat\varg}$ is admissible, the 
Euclidean metric ${\hat\varg}={\mathds 1}$ seems to be the most economical choice, leading to 
standard quantum mechanics with Hermitian observables. Thus, possible physical significances 
of the metric ${\hat\varg}$, or equivalently biorthogonal quantum mechanics, in a unitary system, 
can only be sought in infinite-dimensional systems. 

The introduction of a nontrivial metric operator in Hilbert space emerged independently in the context 
of PT-symmetric quantum mechanics \cite{BBJ,Mostafa}. If a Hamiltonian ${\hat K}$ is symmetric 
under the simultaneous parity-time inversion, then the fact that ${\hat K}$ possesses an anti-linear 
symmetry implies that its eigenvalues can be real. The parity operator ${\hat P}$, however, cannot be 
used as a metric since it is not positive. Nevertheless, associated with such a Hamiltonian is another 
symmetry ${\hat C}$, whose properties resemble those of a charge operator in quantum field theories, 
such that ${\hat\varg}={\hat C}{\hat P}$ can be used as a metric for Hilbert space \cite{BBJ,Mostafa}. 

As a simple example, consider the class of Hamiltonians that are both symmetric \textit{and} 
PT symmetric. The time-reversal operation considered in the literature of PT 
symmetry is usually identified as the operation of complex conjugation. As regard parity 
reversal, in the case of a system modelled on a finite-dimensional Hilbert space there is 
\textit{a priori} no such notion of space reflection, and there is a freedom in the choice of 
the parity operator. A canonical choice, however, is a finite-dimensional analogue of the 
space inversion operator, which is a counter-diagonal matrix whose counter-diagonal 
elements are all unity. With respect to a choice of orthonormal basis $\{|e_n\rangle\}$ we 
can thus write the parity operator ${\hat P}$ in the form: 
\begin{eqnarray}
{\hat P} = \sum_{n} |e_n\rangle\langle e_{N+1-n}| ,
\label{eq:q69}
\end{eqnarray} 
where $N$ is the dimension of the Hilbert space. If the Hamiltonian ${\hat K}$ is symmetric, then 
we have 
\begin{eqnarray}
{\hat K} = \sum_{n,m} \left( \sum_{k,l} K_{kl} u^k_n u^l_m \right) |e_n\rangle\langle e_m| . 
\label{eq:q70}
\end{eqnarray} 
Thus, if we define time reversal to mean complex conjugation, we have 
\begin{eqnarray}
{\hat K}^{PT} = \sum_{n,m} \left( \sum_{k,l} {\bar K}_{kl} 
{\bar u}^k_{N+1-n} {\bar u}^l_{N+1-m} \right) |e_n\rangle\langle e_m| . 
\label{eq:q71}
\end{eqnarray} 
The condition of PT symmetry, however, does not guarantee the reality of the eigenvalues. 
Nevertheless, if, in addition, the eigenstates $\{|\phi_n\rangle\}$ of ${\hat K}$ are also PT 
symmetric, then we have $u^k_n={\bar u}^k_{N+1-n}$. It follows that if a symmetric Hamiltonian 
${\hat K}$ is also PT symmetric, and if the eigenstates of ${\hat K}$ are likewise PT symmetric, 
then $\{K_{nm}\}$ are necessarily real and symmetric (although the matrix elements of ${\hat K}$ 
in an orthonormal basis are not real) so that the eigenvalues of ${\hat K}$ are real. Finally, 
conjugation operation can be defined with the aid of 
\begin{eqnarray}
{\hat C} = \sum_{n} (-1)^n |\phi_n\rangle\langle\chi_{n}| ,
\label{eq:q72}
\end{eqnarray} 
such that ${\hat\varg}={\hat C}{\hat P}$ defines the Hilbert space metric operator. 

One question that arises naturally in this context concerns the combined systems. If one system 
is characterised by the metric operator ${\hat\varg}$, and another by ${\hat\varg}'$, can one 
combine these systems in a meaningful way, and if so, how? Viewed as a system characterised 
by a metric space, the canonical answers to these questions are not immediately apparent; 
however, viewed as a biorthogonal quantum system, the formulation outlined in \S\ref{sec:6} 
provides a canonical way of treating combined systems in this context. In particular, the metric 
operator for the combined system can be constructed from the biorthogonal basis elements of 
the tensor-product space. 

Interests in systems characterised by PT symmetry have increased significantly over the past 
decade due to the observation that PT symmetry can be realised in laboratories by balancing 
gain and loss. Based on the formal equivalence of paraxial approximation to the scalar Hermholtz 
equation and the Schr\"odinger equation (see, e.g., \cite{OS,NM}), first experimental realisations of 
PT-symmetric systems were achieved in optical waveguides \cite{DC2}. Many other experiments 
have subsequently been proposed or realised 
\cite{AM,HS,DC3,SR2,TK,BD,CMB}, 
although it should be added that experiments that have been realised so far 
involve classical systems, where measured quantities do not correspond to eigenvalues of an 
observable acting on states of Hilbert space. 

Quantum mechanically, the implication of the statement of Proposition~\ref{prop:1} on PT symmetry 
is that whether a system is in complete isolation in the sense that all physical observables are 
Hermitian, or whether the system is linked to an environment such that gain and loss are balanced to 
the extent that all eigenmodes are PT symmetric, an observer cannot detect any difference in the 
behaviour of the system. An interesting feature of PT-symmetric systems, however, is that most of 
the model Hamiltonians considered in the literature admit a tuneable parameter (or a set of tuneable 
parameters) such that even though the Hamiltonian ${\hat K}$ is PT symmetric, there are regions in 
parameter space where the eigenstates of ${\hat K}$ are not PT symmetric. In other words, the 
system admits two distinct phases (cf. \cite{BG2}) associated with broken and unbroken PT 
symmetry, and at the transition point the eigenstates of ${\hat K}$ become degenerate (hence 
constitutes an example of an exceptional point). That the 
eigenstates are degenerate implies that they lose the privileged status of being complete; it follows 
from (\ref{eq:q26}) that the operator ${\hat u}$ is not invertible, and consequently the metric operator 
${\hat\varg}$ ceases to exist. Hence an experimental detection of a PT phase transition in a purely 
quantum system modelled on a finite-dimensional Hilbert space will imply that physics beyond 
Hermitian Hamiltonians is not merely an intellectual curiosity but rather is a requirement for the 
description of observed phenomena even in the unitary contexts.

\section{Discussion: towards infinite dimensional systems} 
\label{sec:10}

The foregoing material has been based entirely on finite-dimensional aspects of biorthogonal 
quantum mechanics. It should be noted that already in quantum mechanics based on conventional 
Hermitian 
operators there are subtleties in going from finite to infinite-dimensional Hilbert spaces, and it 
should be intuitively clear that the matter does not improve when considering quantum mechanics 
beyond Hermitian operators. Thus, it will be neither feasible nor realistic to attempt to develop 
a comprehensive account of biorthogonal quantum theory of infinite-dimensional systems here. 
Indeed, the 
following simple example of Young \cite{young} already illustrates how a completeness statement 
of biorthogonal quantum mechanics that holds true in finite dimensions can easily fail in infinite 
dimensions. 

Consider an infinite-dimensional Hilbert space ${\mathcal H}$ and an orthonormal set of basis 
$\{|e_n\rangle\}$ in ${\mathcal H}$. Construct a new set of basis elements $\{|\phi_n\rangle\}$ 
according to the prescription 
\begin{eqnarray}
|\phi_n\rangle = |e_1\rangle + |e_n\rangle
\end{eqnarray}
for $n=2,3,\ldots,\infty$. Evidently, elements of $\{|\phi_n\rangle\}$ are not orthogonal, but the set 
is nonetheless complete since 
\begin{eqnarray}
\lim_{N\to\infty} \frac{1}{N-1} \sum_{n=2}^N |\phi_n\rangle = 
|e_1\rangle + \lim_{N\to\infty} \frac{1}{N-1} \sum_{n=2}^N |e_n\rangle = |e_1\rangle, 
\label{eq:w65}
\end{eqnarray}
on account of the fact that the term orthogonal to $|e_1\rangle$ in the left side of (\ref{eq:w65}) 
decays at the rate $(N-1)^{-1/2}$. It should be evident that the biorthogonal pair of $|\phi_n\rangle$ is 
unique and is given by 
\begin{eqnarray}
|\chi_n\rangle = |e_n\rangle
\end{eqnarray}
for $n=2,3,\ldots,\infty$, so that we have $\langle\chi_n|\phi_m\rangle=\delta_{nm}$. While the 
set $\{|\phi_n\rangle\}$ is complete, its biorthogonal counterpart $\{|\chi_n\rangle\}$ is not---a 
phenomenon that has no analogue in finite dimensions. Thus, if ${\hat K}=\sum_n \kappa_n 
|\phi_n\rangle\langle\chi_n|$ is a Hamiltonian operator acting on the states of ${\mathcal H}$, 
then we can form a linear combination of the eigenstates of ${\hat K}$ according to 
(\ref{eq:w65}) that has a null conjugate state: 
\begin{eqnarray}
\langle{\tilde e}_1|e_1\rangle=0 .
\end{eqnarray}
If we interpret the norm as representing the probability of finding a particle in the system, then 
we have a `no-particle' state $|e_1\rangle$ that nevertheless has nonzero energy expectation 
value, since $\langle{\hat K}\rangle$ in the state $|e_1\rangle$ is formally given by the uniform 
average of the energy eigenvalues, which may be finite or infinite, but will be nonzero. 

Even if a biorthonormal set $(\{|\phi_n\rangle\},\{|\chi_n\rangle\})$ is complete, there can be 
various subtleties arising from the lack of a bounded map that takes an element $|\phi_n\rangle$ 
into $|e_n\rangle$. 
Specifically, suppose that $(\{|\phi_n\rangle\},\{|\chi_n\rangle\})$ is a complete biorthonormal set 
of bases in 
the Hilbert space ${\mathcal H}={\mathcal L}^2$ of square-integrable functions. Then the set 
$\{|\phi_n\rangle\}$ is called a `Fischer-Riesz' basis if (a) for any $|\psi\rangle\in{\mathcal H}$ 
we have $\sum_n|\langle\chi_n|\psi\rangle|^2<\infty$; and (b) if for any sequence $\{c_n\}$ such 
that $\sum_n|c_n|^2<\infty$ there exists a $|\psi\rangle\in{\mathcal H}$ for which 
$\langle\chi_n|\psi\rangle=c_n$. A theorem of Bari \cite{Bari} then shows that: 
(i) $\{|\chi_n\rangle\}$ is a Fischer-Riesz basis if and only if there exists a bounded invertible 
linear operator ${\hat u}^{-1}$ and a complete orthonormal basis elements $\{|e_n\rangle\}$ in 
${\mathcal H}$ such that ${\hat u}^{-1}|\phi_n\rangle=|e_n\rangle$; and that (ii) $\{|\phi_n\rangle\}$ 
is a Fischer-Riesz basis if and only if there exists a positive bounded invertible linear operator 
${\hat\varg}^{-1}$ in ${\mathcal H}$ such that $|\phi_n\rangle={\hat\varg}^{-1}|\chi_n\rangle$. 

In \S\ref{sec:9} we have shown that these results are easily verified in the case of a 
finite-dimensional Hilbert space. In infinite dimensions, on the other hand, a generic complex 
Hamiltonian ${\hat K}$ possessing real eigenvalues often do not admit an invertible bounded metric 
operator ${\hat\varg}$, and this implies that a system described by such a Hamiltonian is intrinsically 
different from that described by a Hermitian Hamiltonian, even if the eigenvalues coincide. There 
is an active research into identifying various implications of the lack of such metric operators in 
various systems \cite{KS,Bagarello,SK,BK,mostafa2}, however, observable effects relating to 
these subtleties have yet to be identified. 

In conclusion, let us summarise the main message of the paper. In the case of quantum systems 
modelled on finite-dimensional Hilbert spaces, provided that an operator possesses real 
eigenvalues and a complete set of eigenstates, it is a viable candidate to represent a physical 
observable, irrespective of whether it is Hermitian in the conventional sense. In particular, there 
seems to be no experiment that one can perform to determine overlap distances between the 
eigenstates in a Hilbert space ${\mathcal H}$, since nonorthogonal eigenstates in ${\mathcal H}$ 
nevertheless correspond to orthogonal states in the projective Hilbert space, in the framework of 
biorthogonal (and unitary) quantum mechanics. The situation, of course, changes if one is 
characterising manifestly open quantum systems lacking unitarity, for which one or more of the 
eigenvalues are not real (see, e.g., \cite{Berry} for a discussion on the determination of the 
Petermann factor $[\langle\chi_n|\phi_n\rangle\langle\phi_n|\chi_n\rangle/\langle\chi_n
|\chi_n\rangle\langle\phi_n|\phi_n\rangle]^{-1}$ in an optical cavity, or \cite{FS} for a discussion 
on the detection of the lack of orthogonality from the statistics of resonance widths). 

Whether the same conclusion concerning the lack of identifiability of the orthogonality of states 
in a unitary theory extends into infinite-dimensional Hilbert spaces remains an open 
question. In this case, the wave function encodes information concerning the configuration of the 
space in which particles exist, in the form of asymptotic boundary conditions. For example, for a 
one-dimensional system, the wave function may be defined on the real line, or along a contour in 
the complex plane (such as the PT-symmetric negative quartic potential \cite{BB}), depending on 
the relevant boundary conditions. Since any such contour can lie along the real axis in a region 
that is experimentally relevant, it is not \textit{a priori} clear whether local measurements 
performed in this region can determine if the wave function should decay along a straight line or 
along a curve at infinities.



\section*{Acknowledgements}

The author acknowledges Eva-Maria Graefe, Bernhard Meister, and Matthew Parry for 
stimulating discussion, and the two anonymous referees for helpful comments.

\bibliographystyle{mdpi}
\makeatletter
\renewcommand\@biblabel[1]{#1. }
\makeatother

\end{document}